\documentclass[journal=jacsat,layout=twocolumn,manuscript=article]{achemso}

\usepackage[utf8]{inputenc}
\usepackage{amsmath}
\usepackage{braket}

\SectionNumbersOn

\title{Chiral Polaritons Based on Achiral Fabry--P\'erot Cavities Using Apparent Circular Dichroism}

\author{Andrew H. Salij}
\affiliation{Department of Chemistry, Northwestern University, 2145 Sheridan Road, Evanston, Illinois 60208, USA}
\author{Randall H. Goldsmith}
\affiliation{Department of Chemistry, University of Wisconsin-Madison, Madison, WI 53706-1322, USA}
\author{Roel Tempelaar}
\affiliation{Department of Chemistry, Northwestern University, 2145 Sheridan Road, Evanston, Illinois 60208, USA}
\email{roel.tempelaar@northwestern.edu}

\begin{tocentry}
\includegraphics[width=\columnwidth]{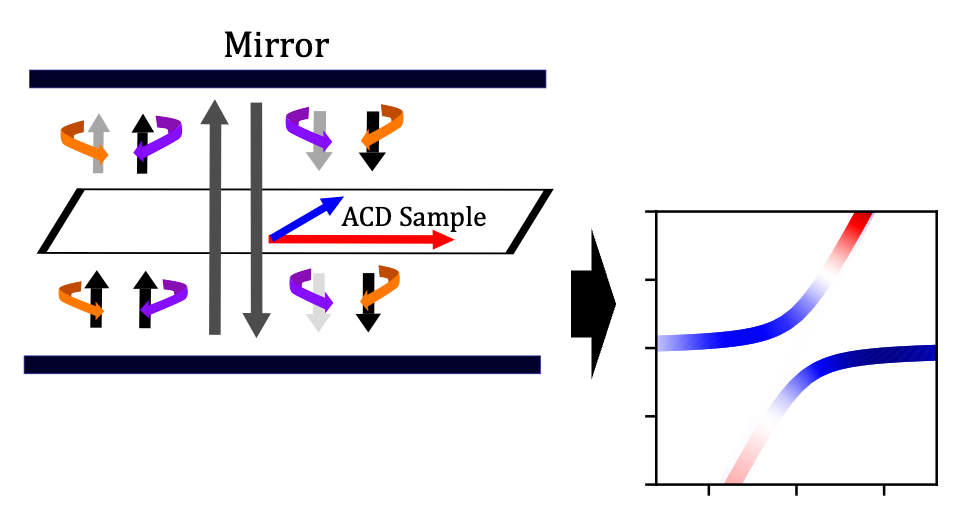}
\end{tocentry}

\newcommand{\vargammafactor}{0.10} 
\newcommand{\varmufigchiral}{10.0~\mathrm{D}}
\newcommand{\vareinf}{8.0} 
\newcommand{\varvol}{4.5~\mathrm{nm}^3}  

\makeatletter
\let\l@addto@macro\relax
\makeatother
\usepackage[fontsize=11pt]{scrextend}

\begin{document}

\maketitle

\begin{abstract}
Realizing polariton states with high levels of chiral dissymmetry offers exciting prospects for quantum information, sensing, and lasing applications. Such dissymmetry must emanate from either the involved optical resonators or the quantum emitters. Here, we theoretically demonstrate how chiral polaritons can be realized by combining (high quality factor) achiral Fabry--P\'erot cavities with samples exhibiting a phenomenon known as ``apparent circular dichroism'' (ACD), which results from an interference between linear birefringence and dichroic interactions. By introducing a quantum electrodynamical theory of ACD, we identify the design rules based on which the dissymmetry of chiral polaritons can be optimized.
\end{abstract}

Keywords: \emph{Strong coupling, polaritons, circular dichroism, thin films, quantum information}

\section{Introduction}\label{sec:intro}

With the rapid developments in the areas of quantum computing, quantum sensing, and quantum communication there is an increased interest in photons as candidate quantum information carriers \cite{stobinska2009perfect, an2009quantum, reiserer2014quantum, northup2014quantum, flamini2018photonic}. Photons are both highly mobile and weakly interacting, as a result of which their quantum state can be transported over long distances, while quantum information can be conveniently stored in their internal spin degree of freedom. Importantly, the manifestation of this spin degree of freedom as circularly-polarized optical polarization allows this information to be transduced to matter by means of chiroptical interactions, allowing photons to be straightforwardly incorporated in quantum networks \cite{wang2016circular, basov2016polaritons, ma2017chiral, gao2017excitonic, lodahl2017chiral, kim2019chiroptical}. Importantly, in addition to chiral selectivity, such implementations require the intrinsically-weak light--matter interactions to be amplified. A viable means to realize this is by optical resonators, through which the strong coupling regime is readily accessible \cite{kavokin2017microcavities}. Within this regime, photons hybridize with excitations of the involved material (quantum emitter), producing polaritons \cite{baranov2018novel, ribeiro2018polariton, yuen2019polariton, herrera2020molecular, keeling2020bose, hertzog2019strong}. Chiral polaritons, where strong coupling is combined with chiral selectivity, would offer the ideal conditions for photon-to-matter quantum transduction, while also being of interest to chiral sensing \cite{wolf2013chirality, heylman2017optical} and lasing.

\begin{figure}[t]
    \centering
    \includegraphics{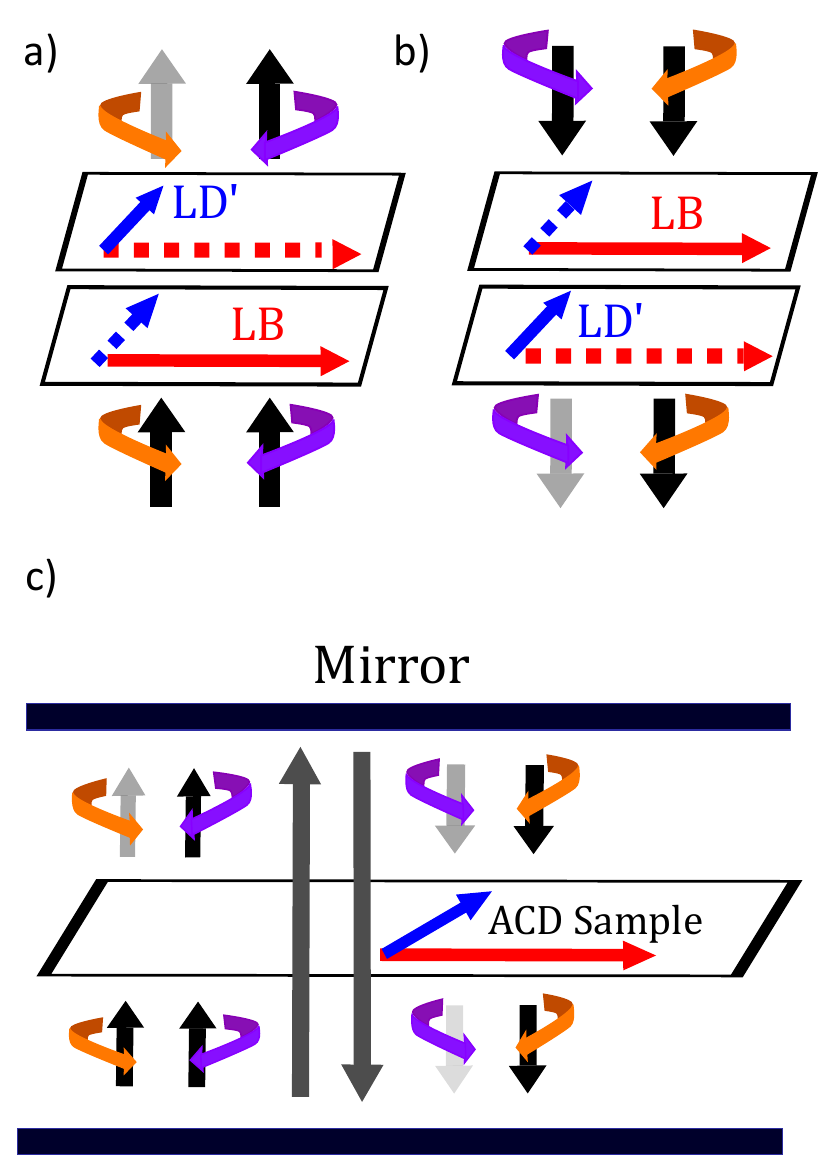}
    \caption{Schematic depiction of ACD resulting from forward (a) and backward (b) propagation, yielding absorption of right-handed (orange) and left-handed (purple) circularly-polarized light, respectively. Also shown is a depiction of the interactions between an ACD sample and an achiral FP cavity (c) involving the selective absoption of the $-$ polarization mode.}
    \label{fig:scheme}
\end{figure}

The chiral selectivity required to produce chiral polaritons should derive from either the quantum emitter or the optical resonators (or both). The latter is an interesting line of inquiry \cite{plum2015chiral, mai2019broadband, feis2020helicity, taradin2021chiral, hubener2021engineering, salij2021microscopic, sun2022polariton, gautier2022planar, voronin2022single}, but involves the challenge of simultaneously optimizing for chiral selectivity \emph{and} for the quality factor. Chiral quantum emitters, on the other hand, have been explored predominantly in the context of chiral plasmonics \cite{zhang2011chiral, du2020chiral, heydari2020analytical, kim2020single, guo2021optical, lin2021chiral}. Perhaps due to the difficulty of overcoming losses involved in plasmonics, recent years have seen a particular interest in polaritons emerging in Fabry--P\'erot (FP) cavities consisting of two parallel mirrors with minimal losses \cite{hood2001characterization}. Achiral FP cavities with high quality factors would be ideally-suited to support chiral polaritons, but impose an interesting constraint on the involved quantum emitter, namely that it not only has a high degree of chiral selectivity \cite{greenfield2021pathways}, but that this selectivity is \emph{inverted} for counter-propagating photons.\bibnote{This oftentimes referred to as ``nonreciprocal'', not to be confused with Lorentz reciprocity, which is obeyed by this phenomenon.} The latter is due to the photonic chirality switching upon reflection at a mirror (i.e., switching from left- to right-handed and \emph{vice versa}), as depicted in Fig.~\ref{fig:scheme} (c) \cite{gil2017polarized}. As such, the quantum emitter must invert its handedness upon plane reflections parallel to the light propagation direction; a phenomenon referred to as two-dimensional (2D) chirality \cite{hecht1994rayleigh, collins2017chirality}. Optically-active samples \cite{fasman2013circular} instead exhibit three-dimensional (3D) chirality associated with point reflections, and as such interact identically with counter-propagating chiral light, as a result of which they do not produce chiral polaritons.

In this Paper, we demonstrate a rare opportunity for realizing quantum emitters with an inverted chiroptical response based on ``apparent circular dichroism'' (ACD). Not unlike FP cavities themselves, ACD as a scientific phenomenon traces back many decades \cite{disch1969apparent}, but in recent years has seen a marked increase in interest \cite{albano2020chiroptical, albano2022reciprocal}. It is a chiroptical response of differential absorption of left-handed and right-handed circularly-polarized light resulting from a macroscopic linear orientation of non-parallel dichroic and birefringent axes of a sample, as depicted in Fig.~\ref{fig:scheme} (a) \cite{disch1969apparent, shindo1985problems}. Within Mueller calculus, where light--matter interactions are described by matrices that manipulate polarization vectors \cite{jensen1978modulation}, the leading contribution to ACD is given by
\begin{equation}
    \mathrm{ACD}(\omega) = \frac{1}{2}(\mathrm{LD}'(\omega)\cdot\mathrm{LB}(\omega)-\mathrm{LD}(\omega)\cdot \mathrm{LB}'(\omega)).
    \label{eq:LDLB}
\end{equation}
Here, LD and LB represent linear dichroism and linear birefringence, respectively, and the prime indicates a $45^\circ$ axis rotation in the plane of polarization. ACD in principle fulfills the 2D chirality requisite for inverted chiroptical response, as shown in Fig.~\ref{fig:scheme} (b). ACD is of second (or higher) order in terms of optical interactions, such that its magnitude increases with sample thickness. As a result, dissymmetry factors, which quantify chiral selectivity, typically exceed those commonly found for optical activity \cite{zinna2020emergent}.

In a recent work, we presented a microscopic theory of ACD wherein the Mueller calculus treatment embodied by Eq.~\ref{eq:LDLB} was combined with a Lorentz oscillator model. This theory allowed identification of the (supra)molecular design rules for optimizing this chiroptical effect directly based on standard electronic structure calculations. This work was motivated by a recent series of systematic experimental studies on the ACD of organic thin films performed by Di Bari and coworkers \cite{albano2017chiroptical, albano2018outstanding, albano2019electronic, zinna2020emergent, albano2022spatially, albano2022aggregation}. In prior years, ACD was mostly considered to be an optical artifact \cite{wu1990method, kuroda2001solid, merten2008vibrational, hirschmann2021treating} arising especially in dye-doped cholesteric liquids \cite{saeva1971induced, saeva1973cholesteric, norden1978vortical, belyakov1979optics, zsila2015apparent}, self-assembled
fibers in solution \cite{wolffs2007macroscopic, hu2021nanoscale}, dyes bound in a linear orientation via a supporting matrix \cite{ritcey1988cholesteric, shindo1990effect}, and nanowires \cite{hu2021nanoscale}. Perhaps for that reason, earlier theoretical works \cite{schellman1987optical, shindo1985on, berova2012comprehensive, gao2020coupling, buffeteau2005vibrational} restricted themselves to macroscopic Mueller or Jones \cite{jones1948new} treatments, although there are some notable early efforts at connecting ACD to microscopic sample properties \cite{disch1969apparent, tunis1970circular}.

The current Paper introduces a quantum electrodynamical theory of ACD allowing us to predict and characterize chiral polaritons arising when ACD samples are embedded in achiral FP cavities. Building on the linear response treatment from our previous work \cite{salij2021theory}, the theory is based on an appropriately-extended Jaynes--Cummings model \cite{jaynes1963comparison}, considering a single quantum emitter in a perfect (lossless) cavity. It is shown that dissymmetries near the theoretical maximum can be achieved even within the single-molecule limit, as the FP cavity allows light to repeatedly interact with the quantum emitter, thereby increasing the optical path length. Generic rules are presented based on which the dissymmetry can be optimized, while particular attention is paid to the utilization of benzo[1,2-$b$:4,5-$b'$]dithiophene-based (BDT-based) oligothiophene as a chiral quantum emitter.

This Paper is organized as follows. In Section \ref{sec:interaction} we introduce chiral interaction terms arising from ACD, which govern the circularly-polarized transition dipoles through which quantum emitters couple to the optical modes of the cavity. This Section also introduces the quantum electrodynamical model of ACD. In Section \ref{sec:polaritons} we present and discuss chiral polariton results. In Section \ref{sec:TSA}, we provide analytical insights into these results by invoking a truncated polaritonic basis set. Design rules are derived from these insights, as summarized in Section \ref{sec:design}. Section \ref{sec:application} proceeds with an application to BDT-based oligothiophene. We conclude in Section \ref{sec:conclusions}.

\section{Chiral Interaction Terms}\label{sec:interaction}

Our quantum electrodynamical theory of ACD is based on a suitably-modified Jaynes--Cummings model \cite{jaynes1963comparison}. Accordingly, the optical polarization is described using the natural basis of orthogonal circularly-polarized modes inside an idealized Fabry--P\'erot (FP) cavity \cite{gutierrez2018polariton}, one of which involves left-handed polarization in the forwards direction and right-handed polarization in the backwards direction (referred to as $\lambda=+$), and \emph{vice versa} for the other (referred to as $\lambda=-$). Accordingly, the light--matter interaction Hamiltonian (within the rotating wave approximation) takes the form
\begin{equation}
    \hat{H}_{\mathrm{int}} = i\sum_n\sum_{\lambda =\pm} A_{0,\lambda}\omega_{n}\tilde{\mu}_{n,\lambda}(\hat{a}_{\lambda}^\dagger \hat{b}_n-\hat{a}_{\lambda} \hat{b}_n^\dagger),
    \label{eq:H_int}
\end{equation}
where $A_{0,\lambda}$ is the vector potential associated with the mode of circular polarization $\lambda$, and where $\hat{a}_\lambda^\dagger$ and $\hat{a}_\lambda$ are the associated photon creation and annihilation operators, respectively. If instead, the light--matter interaction Hamiltonian was expressed in terms of the more commonly-used $x$ and $y$ linearly polarized optical basis, our analysis would not change but the resulting expressions would be less intuitive. In Eq.~\ref{eq:H_int}, $n$ runs over the excited states of the sample, with $\hat{b}_n^\dagger$ and $\hat{b}_n$ as the corresponding creation and annihilation operators, respectively, and with $\hbar\omega_n$ as the associated excited state energy (where the ground state energy is taken to be zero as a reference). The associated transition dipole moment $\tilde{\mu}_{n,\lambda}$ is defined specifically for the $\lambda$ circular polarization, and therefore obeys the same inversion antisymmetry. Notably, achiral as well as 3D chiral samples have $\tilde{\mu}_{n,+} = \tilde{\mu}_{n,-}$, as a result of which there is no chiral selectivity with regard to the circularly-polarized modes of the FP cavity.

In order to derive $\tilde{\mu}_{n,\lambda}$ resulting from ACD we first proceed to consider this phenomenon in terms of linear response theory where light-matter interactions are treated perturbatively. For molecular crystals where intermolecular interactions are negligible, we have previously shown the ACD transition rate to take the form \cite{salij2021theory} 
\begin{align}
\mathrm{ACD}(\omega)&=\frac{1}{2}l^2\xi^2 \omega^2 \label{acd_pert} \\
& \times\sum_{n,m} \mu_n^2\omega_n\mu_m^2\omega_mV_n(\omega) W_m(\omega) \sin(2\beta_{nm}),\nonumber
\end{align}
where $\omega$ is the optical (angular) frequency and $l$ is the sample thickness. Here, $n$ and $m$ run over the excited states of the involved molecule, with $\mu_n^2 = \tilde{\mu}_{n,+}^2+\tilde{\mu}_{n,-}^2$ as the squared total dipole moment of state $n$, and $\beta_{nm}$ as the angle between the transition dipoles associated with states $n$ and $m$. It is assumed that all transition dipoles lie in the plane perpendicular to the light-propagation direction ($xy$ plane), or that any transition dipoles appearing in our analysis have been projected into this plane. In Eq.~\ref{acd_pert}, $\xi \equiv \frac{1}{\hbar c v\epsilon_0\sqrt{\epsilon_\infty}}$ with $v$ being the unit cell volume of the molecular crystal and $\epsilon_\infty$ being its effective high-frequency dielectric constant.\bibnote{Note that in our previous work\cite{salij2021theory} $l$ was included in $\xi$.} Also appearing in Eq.~\ref{acd_pert} are the lineshape functions
\begin{align}
W_n(\omega) &\equiv \frac{\omega_n^2-\omega^2}{(\omega_n^2-\omega^2)^2+\gamma_n^2\omega^2}, \nonumber \\
V_n(\omega) &\equiv \frac{\gamma_n\omega}{(\omega_n^2-\omega^2)^2+\gamma_n^2\omega^2}.
\end{align}
These correspond to the real and imaginary frequency-dependent components of the dielectric susceptibility due to excited state $n$, respectively, and $\gamma_n$ is the associated damping parameter accounting for lineshape broadening.

Eq.~\ref{acd_pert} embodies a Fermi's Golden Rule treatment of ACD, similarly to that of mean (linear) absorption, the latter of which is given by
\begin{equation}
    \bar{A}(\omega)=l\xi\omega\sum_n\mu_n^2\omega_n V_n(\omega).
    \label{eq:abs}
\end{equation}
Alternatively, mean absorption can be expressed as $\bar{A}(\omega)=\frac{1}{2}\left(A_+(\omega)+A_-(\omega)\right)$, with the $\lambda=\pm$ contributions given by
\begin{equation}
    A_\lambda(\omega) \equiv 2l\xi\omega\sum_n\tilde{\mu}_{n,\lambda}^2\omega_n V_n(\omega).
\end{equation}
Key to establishing the form of $\tilde{\mu}_{n,\lambda}$ is that ACD can equivalently be expressed as $\mathrm{ACD}(\omega)=\frac{1}{2}(A_+(\omega)-A_-(\omega))$. By comparison with Eq.~\ref{acd_pert}, it then follows that
\begin{equation}
\tilde{\mu}_{n,\lambda} \equiv \mu_{n}\sqrt{\frac{1}{2}+\frac{1}{2}\tau_\lambda \sigma_n},
\label{eq:mu_tilde}
\end{equation}
where $\tau_\pm = \pm1$ is an indexing variable, and $\sigma_n$ is the ``chiral interaction term'' defined as
\begin{equation}
    \sigma_n \equiv \frac{1}{2} l \xi\omega \sum_{m\neq n} \mu_m^2\omega_m W_m(\omega) \sin(2\beta_{nm}).
    \label{eq:chiral_int_term}
\end{equation}
Here, it can be seen that for a given excited state $n$, other excited states project onto the chiral interaction terms through their real dielectric dispersion, $W_m(\omega)$. As a result, $\tilde{\mu}_{n,\lambda}$ attains an $\omega$ dependence through $\sigma_n$. It is also notable that $\sigma_n$ in principle should be physically bounded by $-1$ and $1$, corresponding to transition dipoles being entirely $-$ or $+$ polarized, respectively. However, $\sigma_n$ depends linearly on $l$, as a result of the second-order Mueller calculus treatment applied in order to arrive at Eq.~\ref{acd_pert}, which could lead to unphysical values of $\sigma_n$, as detailed below.

Having established the chiral transition dipole moments, we will proceed to consider a quantum electrodynamical model of ACD tailored to FP cavities. Accordingly, we replace the free-field optical frequency $\omega$ by the resonance frequency of the FP cavity $\Omega$ set by the mode volume. Here, we will limit ourselves to the lowest-frequency resonance. Furthermore, for the case of a FP cavity, $l$ changes meaning from sample thickness to total optical path length due to the possibility of repeated passes through the sample upon internal cavity reflection.\bibnote{In the case of a bulk material, this would instead be referred as ``penetration depth''.} In practice, this path length will be limited by the cavity quality factor. In this work we will ignore this effect, reserving its inclusion to a follow-up study. Instead, we loosely define the path length as the distance required for electromagnetic intensity to decay to $1/e$ of its original magnitude when (repeatedly) propagating through the sample, meaning the value at which $\bar{A}(\Omega)=1$ (assuming Arhennius-type isotropic absorption). This yields
\begin{equation}
    l = \frac{1}{ \Omega\xi\sum_n \mu_n^2\omega_nV_n(\Omega)},
    \label{eq:penetration_depth}
\end{equation}
where we note that $l$ attains a $\Omega$ dependence due to the dispersion of the sample absorption. Substituting Eq.~\ref{eq:penetration_depth} into Eq.~\ref{eq:chiral_int_term} then yields
\begin{equation}
    \sigma_n = \frac{1}{2}\frac{\sum_{m\neq n} \mu_m^2\omega_m W_m(\Omega)\sin(2\beta_{nm})}{\sum_m\mu_m^2\omega_mV_m(\Omega)}
    \label{chiral_pert}.
\end{equation}
Interestingly, this form of $\sigma_n$ is independent of path length.

It is instructive to consider $\sigma_n$ for the case of two transition dipoles, which is the minimal configuration giving rise to finite ACD, and which eliminates the possibility of \emph{non-inverted} ACD from arising \cite{wolffs2007macroscopic, salij2021theory}. In this case, Eq.~\ref{chiral_pert} simplifies to
\begin{align}
    \sigma_1 &= \frac{1}{2} \frac{W_2(\Omega)/V_2(\Omega)}{1+\Gamma_V\Gamma_{\mu^2}\Gamma_{\omega}}\sin(2\beta_{12}),\nonumber\\
\sigma_2 &= -\frac{1}{2}\frac{W_1(\Omega)/V_1(\Omega)}{1+\Gamma_V^{-1}\Gamma_{\mu^2}^{-1}\Gamma_{\omega}^{-1}}\sin(2\beta_{12}),
\label{sigma_21}
\end{align}
where
\begin{equation}
    \Gamma_V \equiv \frac{V_1(\Omega)}{V_2(\Omega)},\quad \Gamma_{\mu^2} \equiv \frac{\mu_1^2}{\mu_2^2},\quad
    \Gamma_\omega \equiv \frac{\omega_1}{\omega_2}\nonumber.
\end{equation}
We furthermore have that the numerators in Eq.~\ref{sigma_21} can be simplified as $W_n(\Omega)/V_n(\Omega) = \frac{\omega_n^2-\Omega^2}{\gamma_n\Omega}$. Eq.~\ref{sigma_21} elucidates how different parameters impact the magnitude of the chiral interaction terms. Specifically, $|\sigma_{1}|$ and $|\sigma_{2}|$ are both maximized when $\beta_{12} = 45^{\circ}$, which was previously recognized to be the angle of maximum ACD \cite{wolffs2007macroscopic, albano2017chiroptical, salij2021theory}. Interestingly, $|\sigma_1|$ and $|\sigma_2|$ are oppositely impacted by $\Gamma_V$, $\Gamma_{\mu^2}$, and $\Gamma_\omega$, as increasing their values minimizes $|\sigma_1|$ while it maximizes $|\sigma_2|$. It is also noteworthy that $W_n(\Omega)=0$ at $\Omega=\omega_n$, as a result of which $\sigma_1=0$ at $\Omega=\omega_2$ and $\sigma_2=0$ at $\Omega=\omega_1$.

\begin{figure}[t]
    \centering
    \includegraphics{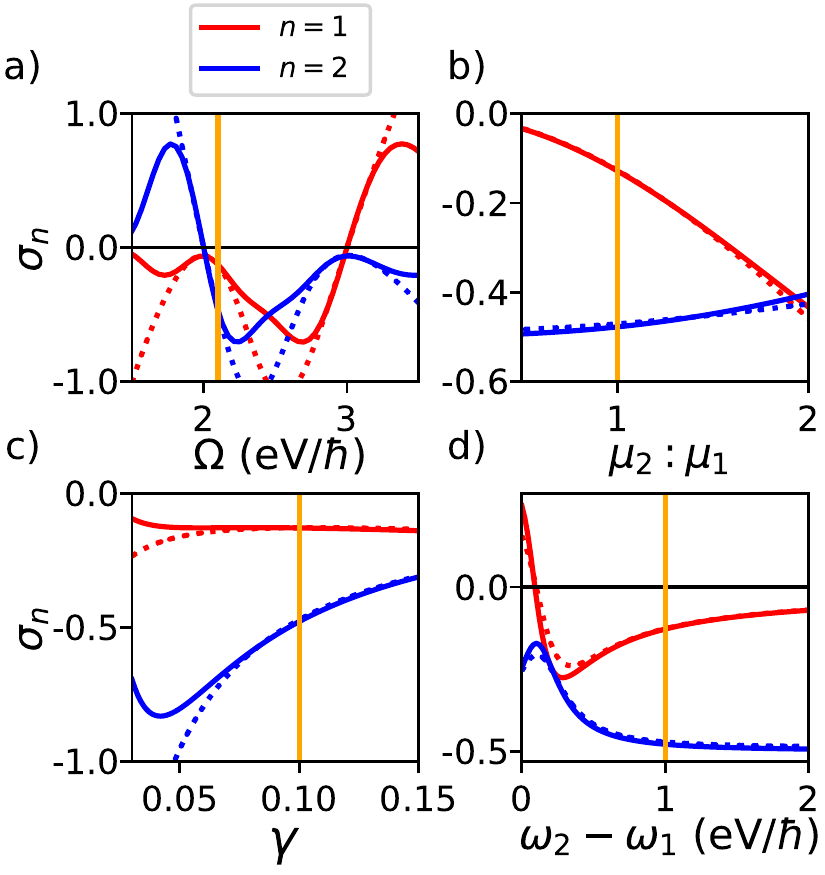}
    \caption{Chiral interaction terms $\sigma_n$ for two transition dipoles as a function of cavity frequency $\Omega$ (a), transition dipole moment ratio $\mu_2:\mu_1$ with $\mu_1$ fixed (b), damping parameter $\gamma$ (c), and transition frequency gap $\omega_2-\omega_1$ with $\omega_1$ fixed (d). Shown are results from a second-order Mueller calculus treatment, using Eq.~\ref{sigma_21} (dotted curve), alongside those from an infinite-order treatment (solid curve; see text), with $\omega_1 = 2.0~\mathrm{eV}/\hbar$ and $\omega_2 = 3.0~\mathrm{eV}/\hbar$ [except for (d)], while the angle between transition dipoles is set to $\beta_{12} = 45^{\circ}$. Unless otherwise noted, $\gamma = \vargammafactor$, $\mu_2:\mu_1 = 1.0$, and $\Omega= 2.1~\mathrm{eV}/\hbar$ in order to be slightly off-resonant with $\omega_1$. Other material parameters are set to $\mu_1 = \varmufigchiral$, $v = \varvol$, and $\epsilon_\infty = \vareinf$. The orange vertical line indicates identical data slices over subplots.}
    \label{fig:chiral_int}
\end{figure}

All of the trends discussed in the above are captured in Fig.~\ref{fig:chiral_int}, which shows $\sigma_1$ and $\sigma_2$ obtained through Eq.~\ref{sigma_21} under sweeps of the various parameters. Here, we have chosen the damping parameter to be proportional to $\omega_n$, such that $\gamma_n = \gamma\omega_n$ for constant $\gamma$, in order to account for the lifetimes of higher-lying states being typically shorter. From Fig.~\ref{fig:chiral_int} (a) it can be seen that $\sigma_1(\Omega-\omega_1)\approx\sigma_2(\omega_2-\Omega)$, which would be a rigorous equality if we assumed $\gamma_1=\gamma_2$. Fig.~\ref{fig:chiral_int} (b) reflects the aforementioned scaling of both chiral interaction terms with $\Gamma_{\mu^2}$. In this and the remaining Panels the chiral interaction terms are depicted at $\Omega\approx\omega_1$, for which $\Gamma_V$ increases with increasing frequency gap, $\omega_2-\omega_1$, and with decreasing $\gamma$. For the overall $\gamma$ dependence of $\vert\sigma_1\vert$, this contribution counteracts that of $W_n(\Omega)/V_n(\Omega)$, rendering $\vert\sigma_1\vert$ nearly constant under variations of $\gamma$, as demonstrated in Fig.~\ref{fig:chiral_int} (c). For $\vert\sigma_2\vert$, on the other hand, both contributions act constructively as a result of which this term decreases with $\gamma$. Lastly, Fig.~\ref{fig:chiral_int} (d) depicts how $\vert\sigma_1\vert$ decreases with increasing $\omega_2-\omega_1$ (through an increase in $\Gamma_V$) with the reverse dependence apparent for $\vert\sigma_2\vert$. These trends break down near $\omega_2-\omega_1 = 0$ at which point we instead find that $\sigma_2 = -\sigma_1$, since all parameters appearing in Eq.~\ref{sigma_21} are identical.

From Fig.~\ref{fig:chiral_int} (a) it becomes obvious that Eq.~\ref{sigma_21} may yield chiral interaction terms that break the physical bounds of $-1$ and $1$. This is particularly so for $\Omega$ off resonance with $\omega_1$ and $\omega_2$, where an increasing path length $l$ causes a failure of the underlying second-order Mueller calculus treatment, as previously mentioned. This failure can be remedied by replacing the second-order Mueller calculus treatment by its infinite-order variant, following previous work by Brown \cite{brown1992unified,brown1999general}. Accordingly, $\frac{1}{2}l$ in Eq.~\ref{eq:chiral_int_term} is replaced by an oscillatory function $B_1(l)/l$ \cite{brown1992unified,brown1999general} which needs to be evaluated numerically (see Supporting Information for details). Results for $\sigma_n$ within this improved treatment are shown alongside those from Eq.~\ref{sigma_21} in Fig.~\ref{fig:chiral_int}, and are indeed found to obey the physical bounds. Notably, for $\Omega$ in resonance with $\omega_1$ and $\omega_2$, results from Eq.~\ref{sigma_21} are seen to be in close agreement with the infinite-order treatment. Importantly, even when physically bounded, the chiral interaction terms are seen to approach $\pm1$ for reasonable parameter values, meaning that levels of chirality near the theoretical maximum can be achieved for ACD under ideal FP cavity confinement. For the remainder of this Paper, we exclusively employ the infinite-order chiral interaction terms.

\section{Chiral Polaritons}\label{sec:polaritons}

We will proceed to evaluate the polaritonic states arising from ACD samples embedded in achiral FP cavities. In doing so, we extend the light--matter interaction Hamiltonian from Eq.~\ref{eq:H_int} within the single-molecule limit in order to include the diagonal photonic and molecular excitation energies, yielding the total Hamiltonian
\begin{align}
    \hat{H} &= \hbar\Omega\sum_{\lambda = \pm} \hat{a}^\dagger_\lambda\hat{a}_\lambda +\sum_{n} \hbar\omega_n\hat{b}^\dagger_n\hat{b}_n \label{hamiltontian_total}\\
    &+i\sum_n\sum_{\lambda =\pm} A_{0,\lambda}\omega_{n}\mu_n\sqrt{\frac{1}{2}+\frac{1}{2}\tau_\lambda \sigma_n}(\hat{a}_{\lambda}^\dagger \hat{b}_n-\hat{a}_{\lambda} \hat{b}_n^\dagger).\nonumber
\end{align}
Here, we introduced the chiral interaction terms through Eq.~\ref{eq:mu_tilde}. The polaritonic eigenstates of this Hamiltonian follow from the time-independent Schr\"odinger equation, $\hat{H}\ket{\Psi^\alpha} = E^\alpha\ket{\Psi^\alpha}$. Within the manifold of single excitations (meaning a single photon or molecular excited state) the eigenstates take the general form
\begin{align}
    \ket{\Psi^\alpha} &= C^\alpha_\mathrm{e}\ket{\psi_\mathrm{e}^\alpha}+C_\gamma^\alpha\ket{\psi_\gamma^\alpha}\nonumber,\\
    \ket{\psi_\mathrm{e}^\alpha} &= \sum_n d_n^\alpha \ket{n},\\
    \ket{\psi_\gamma^\alpha} &= \sum_\lambda d_\lambda^\alpha \ket{\lambda},\nonumber
\end{align}
where we first applied an expansion into the total contributions from the molecular excited states (denoted ``e'') and those from the optical modes (denoted $\gamma$), effectively invoking Hopfield coefficients \cite{hopfield1958theory}, followed by sub-expansions of each. In the above, $\ket{n}\equiv\hat{b}_n^\dagger\ket{0}$ and $\ket{\lambda}\equiv\hat{a}_{\lambda}^\dagger\ket{0}$, with $\ket{0}$ representing the vacuum state without molecular or optical excitations.

In order to characterize the polaritonic eigenstates it proves convenient to define a set of scalar metrics, the first of which is given by
\begin{equation}
    g^\alpha \equiv 2\frac{|d_+^\alpha|^2-|d_-^\alpha|^2}{|d_+^\alpha|^2+|d_-^\alpha|^2}.
    \label{eq:metric_g}
\end{equation}
This metric is analogous to the dissymmetry factor commonly used to characterize chiroptical signals, $g\equiv2\frac{A_+-A_-}{A_+ + A_- }$, and quantifies the anisotropy in the admixture of both chiral modes into the polariton. Note that $g^\alpha$ is bounded as $-2\leq g^\alpha \leq 2$. A second metric is introduced in order to quantify the polaritonic mixing,
\begin{equation}
    \chi^\alpha \equiv 2\vert C_\mathrm{e}^\alpha C_\gamma^\alpha\vert.
    \label{eq:metric_chi}
\end{equation}
This metric is maximized for eigenstates evenly split between molecular and photonic excitations, and is bounded as $0 \leq \chi^\alpha \leq 1$. While $g^\alpha$ and $\chi^\alpha$ quantify the degree of chirality and light--matter hybridization, respectively, combining them yields a single scalar metric quantifying chiral light--matter hybridization. Accordingly, the (chiral) polaritonic dissymmetry factor is defined as
\begin{equation}
    \tilde{g}^\alpha \equiv g^\alpha\chi^\alpha,
    \label{eq:metric_g_tilde}
\end{equation}
which is bounded as $-2\leq \tilde{g}^\alpha \leq 2 $. Maximal $\vert\tilde{g}^\alpha\vert$ implies \emph{both} optimal polaritonic mixing and optimal chirality, whereas a vanishing $\tilde{g}^\alpha$ implies that either polaritonic mixing or chirality (or both) is absent.

\begin{figure}[t]
    \centering
    \includegraphics{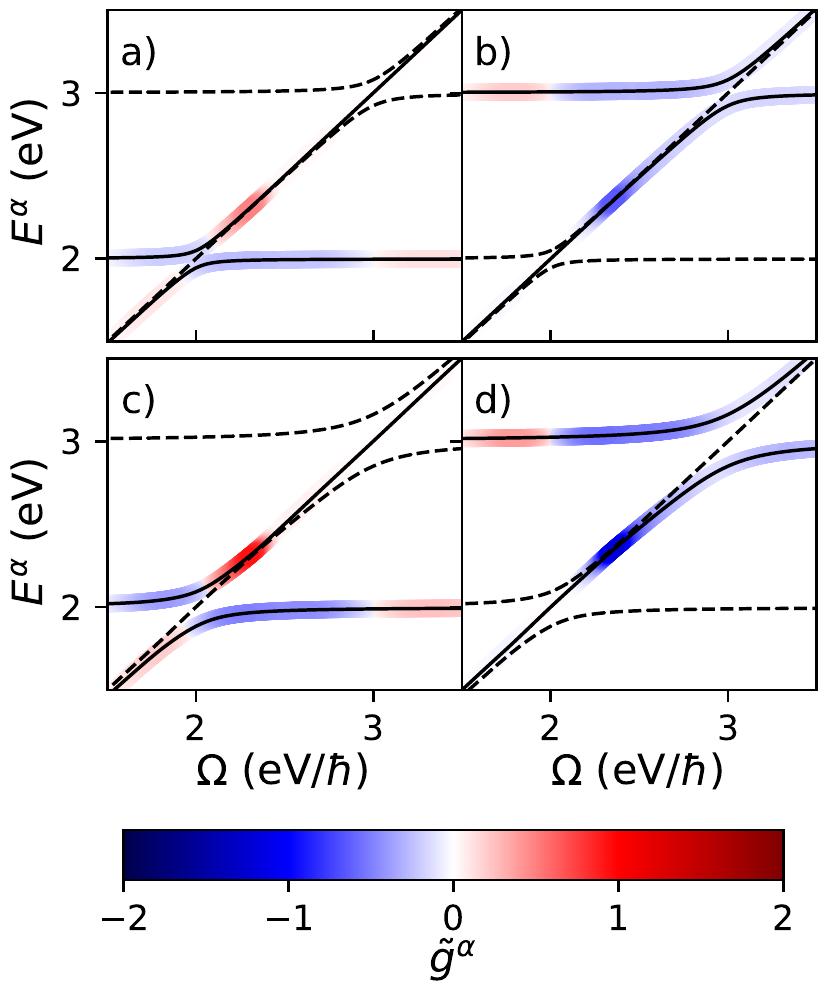}
    \caption{(Chiral) polaritonic dissymmetry $\tilde{g}^\alpha$ as a function of $\Omega$ for the minimal configuration of two transition dipoles. Shown are results for a vector potential $A_{0} = 35~\mathrm{eV}/ec$ (a,b) and for $A_{0} = 70~\mathrm{eV}/ec$ (c,d). Depictions of $\tilde{g}^\alpha$ are separated over two Panels to avoid overlap, and in each Panel correspond to the polariton dispersions indicated by the solid curves (other polariton dispersions are indicated by the dashed curves as a reference). Results are obtained by numerical diagonalization of the Hamiltonian given by Eq.~\ref{hamiltontian_total}, with $\beta_{12} = 45^\circ$, $\omega_1 = 2.0~\mathrm{eV}/\hbar$, $\omega_2 = 3.0~\mathrm{eV}/\hbar$, $\mu_1 = \mu_2 = \varmufigchiral$, $v = \varvol$, and $\epsilon_\infty = \vareinf$.}
    \label{fig:dispersion}
\end{figure}

Shown in Fig.~\ref{fig:dispersion} are polariton dispersions, combined with a representation of the polaritonic dissymmetry $\tilde{g}^\alpha$ as a function of the cavity frequency $\Omega$, assuming an achiral FP cavity by setting $A_{0,+} = A_{0,-} = \frac{1}{\sqrt{2}}A_0$.\bibnote{That is, the total vector potential obeys the Pythagorean equality $A_0^2 = A_{0,+}^2+A_{0,-}^2$ between orthogonal modes and $A_{0,+} = A_{0,-}$.} These results were obtained by solving the time-independent Schr\"odinger equation through numerical diagonalization of the total Hamiltonian given by Eq.~\ref{hamiltontian_total} while representing the molecule by the minimal configuration of two transition dipoles with $\mu_1=\mu_2$ and $\beta_{12} = 45^\circ$. Combined with the two orthogonal photonic states, this yields a total of four eigenstates. These states are depicted in Fig.~\ref{fig:dispersion} for two values of $A_0$ (the largest of which being still amenable to the rotating-wave approximation). The dispersions shown in Fig.~\ref{fig:dispersion} exhibit the known behavior of achiral polaritons, including a Rabi splitting in the regions where $\Omega$ crosses the excited state transitions. Unsurprisingly, this Rabi splitting is seen to increase with $A_0$, replicating the behavior of an achiral Jaynes--Cummings model \cite{jaynes1963comparison}. Importantly, however, there is an undispersed state following the light line in each crossing region.

\begin{figure}[t]
    \centering
    \includegraphics{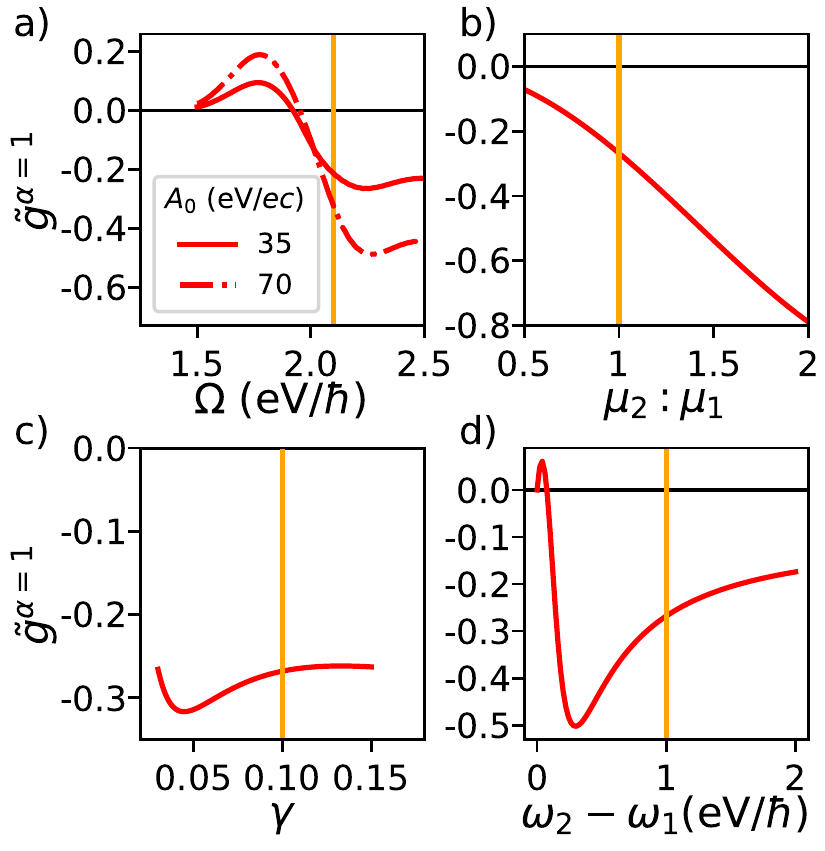}
    \caption{Polaritonic dissymmetry for the lowest-energy polariton, $\tilde{g}^{\alpha=1}$. Shown are results as a function of $\Omega$ (a), $\mu_2:\mu_1$ (b), $\gamma$ (c) and $\omega_2-\omega_1$ (d). Unless noted otherwise, parameters are identical to Fig.~\ref{fig:chiral_int} and with $A_0 = 35~\mathrm{eV}/ec$. The orange vertical line indicates identical data slices over subplots, as in Fig.~\ref{fig:chiral_int}.}
    \label{fig:four}
\end{figure}

In Fig.~\ref{fig:four}, $\tilde{g}^\alpha$ is seen to assume a bisignate profile that changes minimally with increasing $A_0$, apart from an overall increase in amplitude. This suggests that the $A_0$ dependence is primarily confined to the polaritonic mixing $\chi_\alpha$ in Eq.~\ref{eq:metric_g_tilde}, while the (bare) dissymmetry $g^\alpha$ is largely insensitive to $A_0$. Importantly, the polaritonic dissymmetry reaches values of $\tilde{g}^\alpha\sim1.0$, which is a substantial fraction of the theoretically-maximum value. This is particularly remarkable since we are considering a single molecule which within a conventional Mueller calculus treatment of absorption would not have an ACD response (this response being a second-order effect at the minimum) but which is allowed to strongly and repeatedly interact with itself within the FP cavity. Within the resulting sequence of interactions the chirality continuously increases; an effect that is limited by the optical path length, $l$.

Fig.~\ref{fig:four} systematically explores the behavior of $\tilde{g}^{\alpha}$ for the lowest-energy polaritonic eigenstate, $\alpha=1$, which is expected to be the most thermodynamically stable and therefore the most likely candidate for steady-state chiroptical applications. Shown in Fig.~\ref{fig:four} (a) is the $\Omega$ dependence of $\tilde{g}^{\alpha=1}$, emphasizing the bisignate profile previously observed in Fig.~\ref{fig:dispersion}, and further confirming that an increase in $A_0$ primarily acts as an overall rescaling of this quantity governed through $\chi^{\alpha=1}$. Based on Fig.~\ref{fig:four} (b-d) it is tempting to assert that the dependence of $\tilde{g}^{\alpha=1}$ on $\mu_2:\mu_1$, $\gamma$, and $\omega_2-\omega_1$ is instead governed by the bare dissymmetry $g^{\alpha=1}$, seeing the strong similarities with the trends depicted in Fig.~\ref{fig:chiral_int} (b-d), but further analysis is necessary in order to substantiate this.

\section{Three-State Approximation}\label{sec:TSA}

To better understand the computational results shown in Figs.~\ref{fig:dispersion} and \ref{fig:four}, we will proceed with a purely-analytical treatment of the total Hamiltonian given by Eq.~\ref{hamiltontian_total}. This Hamiltonian generally assumes a dimensionality of $(N+2)\times(N+2)$ for a total of $N$ excited states of the sample. Hence, for the minimal configuration of two transition dipoles, one would have to diagonalize a $4\times4$ matrix, which is generally not feasible. However, to an approximation, it is possible to describe the chiral polaritons resulting from Eq.~\ref{hamiltontian_total} by restricting the explicit Hilbert space to a single excited state ($n$), while including the other excited states through their contribution to the chiral interaction term $\sigma_n$. This would be a good approximation provided that excited state $n$ mixes more strongly with the photons than all other states due to having larger coupling elements or due to its transition frequency being closer in resonance with the cavity frequency, $\omega_n \approx \Omega$. Within this ``three-state approximation'' (TSA) the time-independent Schr\"odinger equation takes the form
\begin{align}
    \begin{pmatrix}
    \hbar\Omega & 0 & \Phi_n\sqrt{\frac{1}{2}+\frac{1}{2}\sigma_n}\\
    0 & \hbar\Omega & \Phi_n\sqrt{\frac{1}{2}-\frac{1}{2}\sigma_n}\\
    \Phi_n\sqrt{\frac{1}{2}+\frac{1}{2}\sigma_n} & \Phi_n\sqrt{\frac{1}{2}-\frac{1}{2}\sigma_n} & \hbar\omega_n
    \end{pmatrix}\nonumber\\
    \times\ket{\Psi^\alpha_{(n)}} = E_{(n)}^\alpha\ket{\Psi^\alpha_{(n)}},
    \label{eq:TSA}
\end{align}
where $\Phi_n\equiv |A_0|\omega_n\mu_n$ is the achiral light--matter interaction strength, taken here to be purely real without loss of generality, and where the subscript $(n)$ emphasizes the sample excited state for which the TSA is taken.

The eigenvalue equation given by Eq.~\ref{eq:TSA} is analytically solvable due to the sparsity of the $3\times3$ Hamiltonian matrix. Three solutions are found, one of which consists of purely-photonic contributions,
\begin{equation}
    \ket{\Psi_{(n)}^{\gamma}} = \begin{pmatrix}
    -\sqrt{\frac{1}{2}-\frac{1}{2}\sigma_n}\\
    \sqrt{\frac{1}{2}+\frac{1}{2}\sigma_n}\\
    0
    \end{pmatrix},
\end{equation}
with an associated eigenenergy $E^\gamma_{(n)} = \hbar\Omega$. This explains the undispersed state observed at each crossing in Fig.~\ref{fig:dispersion}. The other two solutions constitute an upper (u) and lower (l) polariton branch, and are given by
\begin{equation}
    \ket{\Psi_{(n)}^\mathrm{u/l}} = \frac{1}{\sqrt{\big(\Pi^\mathrm{u/l}_{(n)}\big)^2+1}}\begin{pmatrix}
\Pi^\mathrm{u/l}_{(n)}\sqrt{\frac{1}{2}+\frac{1}{2}\sigma_n} \\ 
\Pi^\mathrm{u/l}_{(n)}\sqrt{\frac{1}{2}-\frac{1}{2}\sigma_n}\\
1
\end{pmatrix},
\end{equation}
with
\begin{equation}
    \Pi_{(n)}^\mathrm{u/l} \equiv \frac{\Phi_n}{\Delta_n\pm\sqrt{\Delta_n^2+\Phi_n^2}},
\end{equation}
and where $\Delta_n \equiv 2\hbar(\omega_n-\Omega)$ is twice the energetic detuning. The corresponding eigenenergies are given by
\begin{equation}
    E_{(n)}^\mathrm{u/l} = \hbar\frac{\Omega+\omega_n}{2}\pm\sqrt{\Delta_n^2+\Phi_n^2}.
\end{equation}

Substituting the above eigensolutions into the (bare) dissymmetry factor defined in Eq.~\ref{eq:metric_g} yields (see Supporting Information for details)
\begin{equation}
    g_{(n)}^\mathrm{u/l} = 2\sigma_n.
    \label{eq:g_n}
\end{equation}
Interestingly, within the second-order Mueller calculus treatment this dissymmetry factor is independent of $A_0$, cf.~Eq.~\ref{chiral_pert}, with a (weak) dependence only being possible through higher-order effects contained in oscillatory function $B_1(l)/l$. This substantiates our observations in Figs.~\ref{fig:dispersion} and \ref{fig:four} that the effect of $A_0$ is largely manifested in the polaritonic mixing. Within the TSA, this mixing is obtained by substituting the above eigensolutions in Eq.~\ref{eq:metric_chi}, yielding (see Supporting Information for details)
\begin{equation}
    \chi_{(n)}^\mathrm{u/l} = \frac{\Phi_n}{\sqrt{\Delta_n^2+\Phi_n^2}}.
    \label{eq:chi_n}
\end{equation}
As expected, this mixing is maximized to $1$, corresponding to a perfect split between photonic and electronic states at resonance, $\Delta_n=0$, and decreases with increasing $\Delta_n$ or decreasing $\Phi_n$.

\begin{figure}
    \centering
    \includegraphics{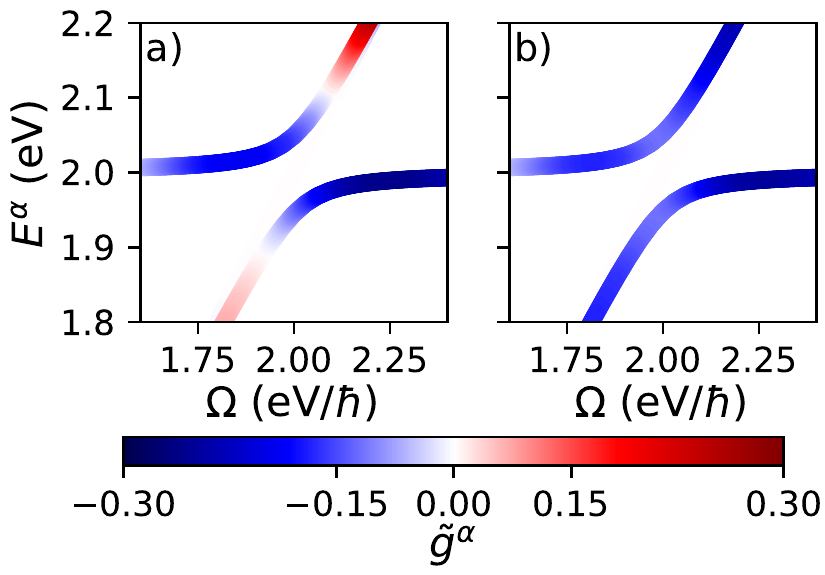}
    \caption{Reproduction of Fig.~\ref{fig:dispersion} with $A_0=35~\mathrm{eV}/ec$ (a) and comparative results within the TSA (b), explicitly including only the lowest-energy excited state ($n=1$). Note the change in colormap scaling and boundaries for easier comparison.}
    \label{fig:one_sigma_vs_two}
\end{figure}

In order to assess the accuracy of the TSA, we compare in Fig.~\ref{fig:one_sigma_vs_two} results against the numerical solutions of the full Hilbert space for the $A_0=35~\mathrm{eV}/ec$ case previously considered in Fig.~\ref{fig:dispersion}. Shown are the polariton dispersions and $\tilde{g}^\alpha$ as a function of $\Omega$. The TSA is taken for $n=1$. The resulting dispersions are indistinguishable from those predicted by the full Hilbert space, indicating that mixing due to $n=2$ has a negligible impact on $E_{(n=1)}^\mathrm{u/l}$. In contrast, discrepancies are observed for the polaritonic dissymmetry factors. Within the TSA the upper and lower polariton pair have identical chirality, $g^\mathrm{u}_{(n=1)} = g^\mathrm{l}_{(n=1)}$, while the overall polaritonic dissymmetry is seen to be monosignate. Within the full Hilbert space, $g^\mathrm{u}_{(n=1)} = g^\mathrm{l}_{(n=1)}$ is violated, with discrepancies occurring exclusively for polaritonic states close to the light line (for which $E_{(n=1)}^\mathrm{u/l}\approx\Omega$), giving rise to the bisignate profile of $\tilde{g}^\alpha$ obtained within the full Hilbert space. This behavior is rationalized by the two excited states of the sample coupling only indirectly to one another through the photonic states. When close to the light line polaritonic states contain a predominant photonic component that couples directly to both excited states, as a result of which $g_{(n)}^\mathrm{u/l} \neq \sigma_n$, whereas when separated from the light line polaritons consist predominantly of a single excited state with only a minor photonic component, and thus weak mixing of the other excited state, as a result of which the TSA is accurate. Within these spectral regions $g_{(n)}^\mathrm{u/l} = \sigma_n$ offers a direct manner by which one can tune the polaritonic chirality.

With the analytical insights offered by the TSA we now revisit Fig.~\ref{fig:four}. For the lowest-energy polaritonic state depicted here, the regime of validity of the TSA is where $\Omega>\omega_1$ (away from the light line) while $\Omega$ still being sufficiently separated from $\omega_2$. Within this regime, $\tilde{g}^{\alpha=1}\approx\chi^{\alpha=1}\sigma_1$, and the profile of $\tilde{g}^{\alpha=1}$ as a function of $\Omega$ can be rationalized based on the dispersion of $\sigma_1$ in Fig.~\ref{fig:chiral_int} (a), while appreciating that $\chi^{\alpha=1}$ monotonically decreases away from the resonance $\Omega\approx\omega_1$. Outside the TSA regime, the sign change of $\tilde{g}^{\alpha=1}$ (giving rise to the bisignate profile) is a direct consequence of a mixing of the $n=2$ excited state. The TSA analysis furthermore confirms that the $A_0$ dependence is confined to $\chi^{\alpha=1}$ whereas the $\mu_2:\mu_1$, $\gamma$, and $\omega_2-\omega_1$ dependence is confined to $g^{\alpha=1}$. In particular, for the results shown in Fig.~\ref{fig:four} (b-d) we have $\chi^{\alpha=1}\approx1$ (due to $\Omega\approx\omega_1$) as a result of which $\tilde{g}^{\alpha=1}\approx2\sigma_1$, which is readily verified through a comparison with Fig.~\ref{fig:chiral_int} (b-d).

\section{Design Rules for Chiral Polaritons}\label{sec:design}

Through both our numerical and approximate analytical treatments of the quantum electrodynamical theory of ACD, we have arrived at a set of design rules for optimizing the chiral selectivity of polaritons, embodied by $\tilde{g}^\alpha$. These design rules can be summarized as follows.
\begin{enumerate}
    \item As with polaritonic states in general, the frequency of the cavity mode should be approximately resonant with that of some transition of the quantum emitter, $\Omega\approx\omega_n$ [see Eq.~\ref{eq:chi_n}].
    \item Other quantum emitter transition frequencies ($\omega_m$) must be sufficiently close to $\Omega$ for the ACD interactions to be relevant, while being sufficiently separated from $\omega_n$ to have well-resolved polariton states [see Fig.~\ref{fig:four} (d)].
    \item Dipole moments of those other transitions are preferably large compared to that of the resonant transition, $\mu_m\ll\mu_n$, while their mutual angle approaches 45$^\circ$ [see Eqs.~\ref{sigma_21} and \ref{eq:g_n}].
    \item Lastly, there are considerations regarding energetic stability (not studied in the present work), which suggests that the resonant transition preferably involves the lowest-energy quantum emitter excited state, rendering the resulting polariton optimally stable against relaxation pathways.
\end{enumerate}

\section{Application to BDT-Based Oligothiophene}\label{sec:application}

Having in place the design rules for optimizing chiral polaritons based on ACD, we now turn our attention to BDT-based oligothiophene. Thin films composed of this molecule were the specific focus of our previous work introducing a microscopic treatment of ACD within linear response theory \cite{salij2021theory}. Importantly, the intermolecular electronic interactions were found to be weak for these films \cite{salij2021theory}, as a result of which our microscopic treatment was directly applicable. Excellent agreement was found for linear absorption and ACD spectra against experimental results \cite{albano2017chiroptical} upon including three electronic transitions coupled to a high-frequency intramolecular vibration, and parametrized based on electronic structure calculations and spectral fitting. We will proceed to theoretically predict the chiral polaritons that would arise when BDT-based oligothiophene serves as a quantum emitter in an achiral FP cavity. As in the previous Sections, we describe this setup within the limit of a single molecule through application of the modified Jaynes--Cummings model incorporating the quantum-electrodynamical theory of ACD.

\begin{figure}[t]
    \centering
    \includegraphics{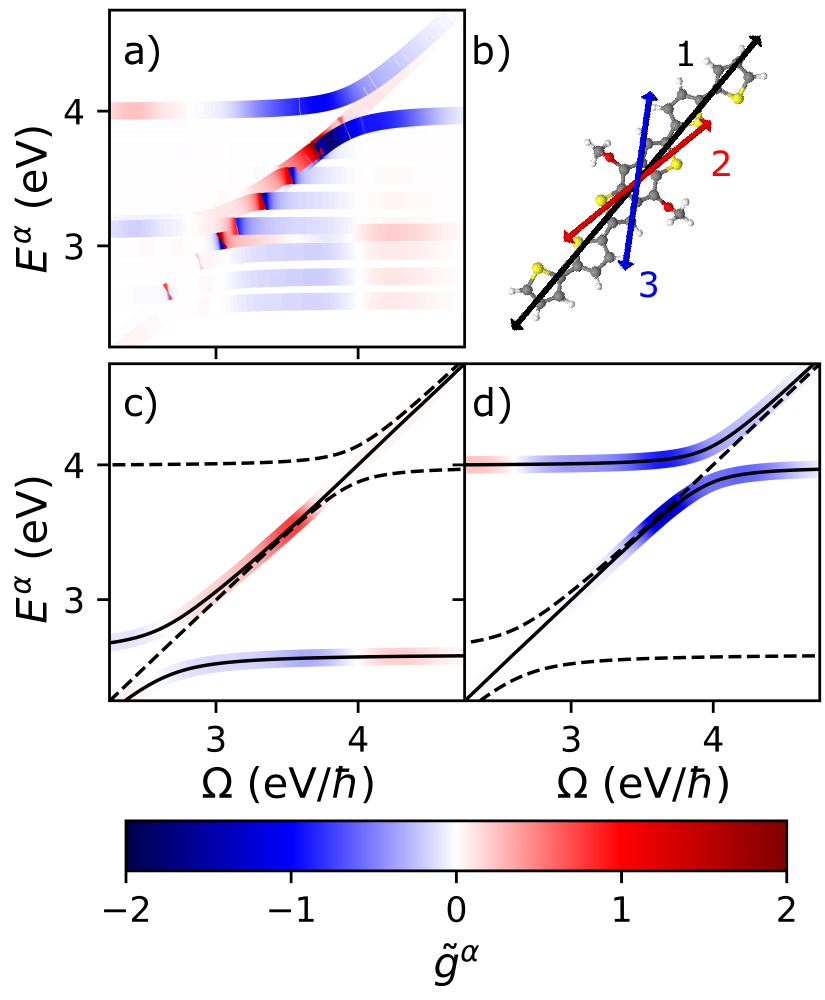}
    \caption{Polaritonic dissymmetry $\tilde{g}^\alpha$ for BDT-based oligothiophene as predicted by our theory (a). Overlapping states drawn in order of $|\tilde{g}^\alpha|$ such that states with greatest $|\tilde{g}^\alpha|$ are visible. The model of the BDT-based oligothiophene was adopted from our previous work \cite{salij2021theory} (see text for details). Molecular structure shown in (b) with transition dipoles indicated. Also shown are results without the vibrational modes and only including the first and third electronic transitions (c, d).}
    \label{fig:di_bari}
\end{figure}

Fig.~\ref{fig:di_bari} (a) shows the calculated polaritonic dissymmetry and dispersions. We refer to our previous work \cite{salij2021theory} for the applicable model parameters, but note that we adjusted the unit cell volume to $v=4.46$~nm$^3$, in order to more adequately account for the oligothiophene side chains, and the high-frequency dielectric constant to $\epsilon_\infty=8.0$, to better agree with relevant experimental high-frequency dielectric constants \cite{okutan2007dielectric}. As previously discussed \cite{salij2021theory}, ACD in BDT-based oligothiophene is due to interactions of a strong electronic transition at $\omega_1=2.6$~eV (with $\mu_1=13.7$~D) interfering with two weaker transitions at $\omega_2=3.1$~eV (with $\mu_2=6.7$~D) and $\omega_3=4.0$~eV (with $\mu_2=6.4$~D) under angles $\beta_{12}=-10.5^\circ$ and $\beta_{13}=31.5^\circ$. The lowest transition is furthermore coupled to a vibrational mode with an energy of 185~meV. The resulting manifold of vibronic excited states gives rise to a large number of pairwise interactions between transitions that contribute to the chiral interaction terms, and thus a large number of chiral polaritons. This is borne out in  Fig.~\ref{fig:di_bari} (a) where a panoply of chiral polaritons can be seen. What is immediately obvious is that very large polaritonic anisotropies are reached, with values of $\vert\tilde{g}^\alpha\vert$ up to $\sim1.8$, which forms a stark contrast with bare BDT-based oligotiophene thin films that reach typical dissymmetry values of $\sim 0.02$\cite{albano2017chiroptical}. Similarly to Section \ref{sec:polaritons}, the significant gain in dissymmetry upon polariton formation is due to the molecule being able to strongly and repeatedly interact with itself within the FP cavity.

In spite of the large number of states, the regions of large $\vert\tilde{g}^\alpha\vert$ can largely be understood based on a simplified model including the electronic transition at $4.0$~eV interacting with the electronic transition at 2.6~eV (and without the inclusion of vibronic coupling). This is borne out in Figs.~\ref{fig:di_bari} (b) and (c), where we reproduce the salient features shown in Fig.~\ref{fig:di_bari} (a) by restricting the model to this pair of transitions. Both the dispersions and polaritonic anisotropies resulting from this reduced model can in turn be understood based on our analysis of the two-dipole system in Fig.~\ref{fig:dispersion}. Moreover, the involved transitions can be shown to satisfy the majority of the design rules presented in the previous Section, especially with the cavity frequency in resonance with the molecular transition at 4.0~eV. In that case, $\tilde{g}^\alpha$ benefits from the comparatively larger dipole moment of the transition at 2.6~eV, which adds to the favorable inter-dipolar angle of $31.5^\circ$\cite{salij2021theory}. A potential pitfall would be that the resonant molecular transition is the higest in energy rather than the lowest, and a reversal of these excited state properties is likely to yield chiral polaritons with an improved energetic stability. Regardless, these results offer encouraging prospects for the practical implementation of chiral polaritons based on existing ACD samples.

\section{Conclusions and Outlook}\label{sec:conclusions}

We have predicted and characterized chiral polariton states arising when ACD samples are embedded in achiral FP cavities, based on a suitably-modified Jaynes--Cummings model incorporating a quantum electrodynamical theory of ACD. Our results suggest the feasibility of employing ACD for the engineering of polaritons with strong chiral selectivity, with dissymmetry factors approaching their theoretical maxima even when taking the quantum emitter in the single-molecule limit. This is due to ACD being of second or higher order in terms of the light--matter interaction, allowing the exceptionally-high dissymmetry factors to be realized as the quantum emitter interacts with itself through the cavity. Moreover, the inverted chiroptical response originating from ACD proves compatible with achiral FP cavities. As such, chiral polariton engineering efforts can optimize exclusively for cavity quality factor, without having to additionally achieve chiral selectivity of the cavity. We applied our theory to BDT-based oligothiophene, which was previously experimentally \cite{albano2017chiroptical} and theoretically \cite{salij2021theory} scrutinized for its pronounced ACD response, and provided indications that high dissymmetries are attainable for experimentally realizable samples. Future efforts will be directed towards understanding collective effects arising from multiple optically-confined ACD quantum emitters as well as the role of the cavity quality factor. As to the former, it will be interesting to see how a helical stacking of molecules, giving rise to a 3D form of ACD with \emph{non-inverted} chiroptical selection rules \cite{salij2021theory}, will inhibit the formation of chiral polaritons. Through these efforts, we hope to contribute to the realization of chiral polaritonic phenomena in order to open up new technological opportunities.

\section{Acknowledgements}

This work was supported as part of the Center for Molecular Quantum Transduction (CMQT), an Energy Frontier Research Center funded by the U.S. Department of Energy, Office of Science, Basic Energy Sciences under Award No.~DE-SC0021314. 

\section{Supporting Information}

Supporting Information includes an infinite-order Mueller calculus treatment, a justification of why mirrors may be treated as reciprocal boundaries of a continuous medium, and derivations of the interaction Hamiltonian, ACD, and the polaritonic characteristics.

\bibliography{biblio}

\begin{mcitethebibliography}{84}
\bibitem[Not()]{Note-1}
This oftentimes referred to as ``nonreciprocal'', not to be confused with
  Lorentz reciprocity, which is obeyed by this phenomenon.\relax
\mciteBstWouldAddEndPunctfalse
\mciteSetBstMidEndSepPunct{\mcitedefaultmidpunct}
{}{\mcitedefaultseppunct}\relax
\EndOfBibitem
\bibitem[Not()]{Note-3}
In the case of a bulk material, this would instead be referred as ``penetration
  depth''.\relax
\mciteBstWouldAddEndPunctfalse
\mciteSetBstMidEndSepPunct{\mcitedefaultmidpunct}
{}{\mcitedefaultseppunct}\relax
\EndOfBibitem
\bibitem[Not()]{Note-4}
That is, the total vector potential obeys the Pythagorean equality $A_0^2 =
  A_{0,+}^2+A_{0,-}^2$ between orthogonal modes and $A_{0,+} = A_{0,-}$.\relax
\mciteBstWouldAddEndPunctfalse
\mciteSetBstMidEndSepPunct{\mcitedefaultmidpunct}
{}{\mcitedefaultseppunct}\relax
\EndOfBibitem
\end{mcitethebibliography}

\end{document}